\documentclass[journal]{IEEEtran}
\usepackage{subfigure}
\usepackage{color}
\usepackage{cite}
\usepackage{amsmath}
\usepackage{cite}
\usepackage{graphicx}
\usepackage{epstopdf}
\usepackage{amsfonts,amsmath,amssymb}
\usepackage{graphicx}
\usepackage{url}
\usepackage{bm}
\usepackage{bbm}
\usepackage{subfigure}
\usepackage{stfloats}
\usepackage{color}

\usepackage{cite}

\DeclareMathOperator*{\argmax}{argmax}

\newcommand{\Hnull}{\mathcal{H}_0}
\newcommand{\Halt}{\mathcal{H}_1}

\newcommand{\Honull}{\mathcal{{D}}_0}
\newcommand{\Hoalt}{\mathcal{{D}}_1}

\usepackage{cite,graphicx,amsmath,amssymb,cite,algorithm}

\newtheorem{corollary}{\textbf{Corollary}}
\newtheorem{theorem}{\textbf{Theorem}}
\newtheorem{proposition}{\textbf{Proposition}}

\begin{document}

\title{Covert Wireless Communications with Channel Inversion Power Control in Rayleigh Fading}
\author{{Jinsong Hu,~\IEEEmembership{Student Member,~IEEE,} Shihao Yan,~\IEEEmembership{Member,~IEEE,} Xiangyun Zhou,~\IEEEmembership{Senior Member,~IEEE,} \\ Feng Shu,~\IEEEmembership{Member,~IEEE,} and Jun Li,~\IEEEmembership{Senior Member,~IEEE}}

\thanks{J. Hu, F. Shu, and J. Li are with the School of Electronic and Optical Engineering, Nanjing University of Science and Technology, Nanjing, China. (Emails: \{jinsong\_hu, shufeng, jun.li\}@njust.edu.cn). F. Shu is also with the College of Computer and Information Sciences, Fujian Agriculture and Forestry University, Fuzhou, China.

S. Yan is with the School of Engineering, Macquarie University, Sydney, NSW, Australia (Email: shihao.yan@mq.edu.au).

X. Zhou is with the Research School of Engineering, Australian National University, Canberra, ACT, Australia (Email:xiangyun.zhou@anu.edu.au).
}
}



\maketitle

\vspace{-2cm}

\begin{abstract}
In this work, we adopt channel inversion power control (CIPC) to achieve covert communications aided by a full-duplex receiver. Specifically, the transmitter varies the power and phase of transmitted signals as per the channel to the receiver, such that the receiver can decode these signals without knowing the channel state information. This eliminates the required feedback from the transmitter to the receiver, which aids hiding the transmitter from a warden. The truncated CIPC and conventional CIPC schemes are proposed and examined, where for truncated CIPC covert transmission ceases when the channel quality from the transmitter to the receiver is low, while for conventional CIPC covert transmission always occurs regardless of this channel quality. We examine their performance in terms of the achieved effective covert throughput (ECT), which quantifies the amount of information that the transmitter can reliably convey to the receiver, subject to the constraint that the warden's detection error probability is no less than some specific value. Our examination shows that the truncated CIPC scheme can outperform the conventional CIPC scheme due to this constraint.
\end{abstract}

\vspace{-0.5cm}



\section{Introduction}
%

With the ever-increasing use of the Internet of Things (IoT), various types of small devices are becoming part of the wireless connected world, whose overall goal is to improve the quality of our daily life. In a wide range of application scenarios, the security of IoT is a critical issue. For example, in health-care systems, some wireless sensors collect patients' health information such as heart rate and blood pressure. This type of information is private and highly confidential and hence a secure transmission is of a high demand. However, due to broadcast nature of the wireless medium, the transmission in IoT can be easily detected or eavesdropped on by unauthorized users \cite{Mukherjee2015Physical,Xu2016Security}.


Traditional security techniques offer protection against eavesdropping through encryption \cite{Menezes1996Handbook,Talbot2007Complexity}, guaranteeing the integrity of messages over the air. However, it has been shown in the recent years that even the most robust encryption techniques can be defeated by a powerful adversary (e.g., a quantum computer) \cite{Rich2014full}. Meanwhile, physical-layer security, on the other hand, exploits the dynamic characteristics of the wireless medium to preserve the confidentiality of the transmitted information in wireless networks \cite{zhaonan2016Anti,yan2016artificial,hu2017artificial,shu2018secure}. We note that both the conventional encryption and physical layer security techniques cannot provide protection against the detection of a transmission in the first place, which may disclose a user's critical information (e.g., exposing a user's location information). As such, hiding a wireless transmission in the first place is widely required in some IoT applications, which is also explicitly desired by government and
military bodies. Against this background, covert communications (also termed low probability of detection communications) are emerging as new and cutting-edge wireless communication security techniques, which aim to enable a wireless transmission between two users while guaranteeing a negligible probability of detecting this transmission at a warden \cite{bash2013limits,bash2015hiding,Boulat2016Time,bloch2016covert,goeckel2016covert,wang2016fundamental}.

In the literature, the fundamental limit of covert communications over additive white Gaussian noise (AWGN) channels has been studied in \cite{bash2013limits}. It is proved that $\mathcal{O}(\sqrt{n})$ bits of information can be transmitted to a legitimate receiver reliably and covertly in $n$ channel uses as $n \rightarrow \infty$. Following \cite{bash2013limits}, covert communications have been studied in a few scenarios. For example, covert communications can be achieved when the warden has uncertainty about the receiver noise power~\cite{lee2015achieving}. In \cite{Sobers2017Covert}, the authors adopted a friendly jammer, which generates artificial noise to create uncertainty at the warden Willie, in order to help achieve covert communications. The authors of \cite{Biao2018Poisson} considered covert communications with a poisson field of interferers, in which it was proved that the density and the transmit power of the interferers do not affect the covert communication performance when the network stays in the interference-limited regime. The covert communication with interference uncertainty from non-cooperative transmitters is studied in \cite{Xidian2018ICC}.
The effect of finite blocklength (i.e., short delay constraints) over AWGN channels on covert communications was examined in \cite{ShihaoYan2017Covert}, which proves that the effective throughput of covert communications is maximized when all available channel uses are utilized. A covert communication system under block fading channels was examined in \cite{Shahzad2017Covert,Shihao2018Delay}, where transceivers have uncertainty on the related channel state information. Covert communications in the context of relay networks was examined in \cite{Jinsong2017GLOBECOM,Hu2018covertrelay}, which shows that a relay can transmit confidential information to the corresponding destination covertly on top of forwarding the source's message.
In \cite{yan2018gaussian}, the optimality of Gaussian signalling was examined in the context of covert communications with two different constraints, where Gaussian signalling was proved to be optimal for one covert communication constraint, but not optimal the other one.

The aforementioned works in the literature mainly focused on how to hide the wireless transmission action (not the transmitter itself), since some information that can indicate the existence of the transmitter was assumed \emph{a priori} known. For example, in \cite{lee2015achieving,BiaoHe2017on,Shahzad2017Covert,Jinsong2017GLOBECOM,ShihaoYan2017Covert} it is assumed that the instantaneous wireless channels from the transmitter to the warden are known by the warden, which means that the warden knows the existence of the transmitter and is to detect whether a wireless transmission occurs. We note that the ultimate goal of covert communications is to achieve a shadow wireless network \cite{bash2015hiding}, in which the transmitter itself should be hidden from the warden. This is due to the fact that the exposure of a user's location information will cause severe negative impact in some applications. For example, when IoT is adopted in battlefields for communication, location information of soldiers or headquarters is extremely confidential, since exposure of this information may lead to fatal attack on the soldiers or headquarters. As such, we aim at an initial step towards this ultimate goal of covert communications by removing some strong assumptions that reveal the existence of the transmitter \emph{a priori}. In this work, we adopt the channel inversion power control (CIPC) at the transmitter (Alice) to achieve covert communications, in which the transmitter varies the power and phase of its transmitted signals as per the channel from itself to the receiver (Bob) which as a known base station in order to keep the signal power at the receiver equal to a certain constant value. Such approach has the benefit of removing the requirement that the receiver has to know the channel state information (that requires the feedback from the transmitter) and thus hiding the existence of the transmitter from warden (Willie) before any covert information transmission.  In a standard (non-covert) communication system, channel inversion power control can significantly decreases the outage probability \cite{Steven2007CIPC}. It is well known that the truncated CIPC scheme is more general, which includes the conventional CIPC scheme as a special case with the maximum value of the transmit power approaches the infinity and can achieve the minimum value of the outage probability. However, when considering covert communications, conventional CIPC scheme might make it easy for the signal to be observed at Willie and hence its utility is less clear, which motivates us to consider the these two CIPC scheme and compare their performance in this work. Moreover, the IoT device (e.g., Alice) is usually equipped with a single antenna each and could not cooperation in the typical application scenario of IoT \cite{wanghuiming2018D2D}. As such, the receiver Bob as more powerful base station can use an additional antenna to transmit artificial noise (AN) for further degrading the detection performance of Willie.

Our main specific contributions are summarized as below.


\begin{itemize}
\item Considering practical Rayleigh fading channels in wireless networks, in this work we adopt the CIPC at the transmitter to achieve covert communications, in which the transmitter varies its transmit power as per the channel from itself to the receiver, such that the received power of the covert information signal is a fixed value $Q$. With channel reciprocity, this power control strategy does not require the transmitter to transmit pilot signals for channel estimation or to feed back the estimated channel to the receiver before covert communications, which may announce the existence of the transmitter before a covert transmission. As such, this power control can potentially aid hiding the transmitter and thus we adopt it in two CIPC schemes to achieve covert communications in this work, where its performance is thoroughly examined.

\item We first consider the truncated CIPC scheme, where the covert transmission ceases when the quality of the channel from the transmitter to the receiver is lower than some specific value (determined by the maximum transmit power constraint). We analyze the detection performance at the warden, based on which we determine effective covert throughput (ECT) that quantifies the amount of information that can be conveyed from the transmitter to the receiver subject to the warden's detection error probability being no less than some specific value. Specifically, the detection error probability at the warden is derived in a closed-form expression, based on which the optimal detection threshold is analytically achieved. The generated AN offers the capability of hiding the transmitter. Our examination shows that the increase in the maximum transmit power of AN at the full-duplex receiver may not continuously improve ECT. This is due to that the transmitted AN causes self-interference at the receiver and the transmitter is subject to a maximum power constraint.

\item We also consider the conventional CIPC scheme as a benchmark to examine the performance limit of CIPC in an asymptotic scenario, where there is no maximum power constraint at the transmitter and the covert transmission always occurs regardless of the channel quality. Solid performance analysis on this scheme has conducted, since mathematically the conventional CIPC scheme cannot be a special case of the truncated CIPC scheme. For the conventional CIPC scheme, our analysis shows that the value of $Q$ can be varied to counteract the impact of the transmit power of AN on the warden's detection performance and thus the achieved ECT approaches an analytical derived upper bound as the maximum transmit power of AN tends to infinity.
\end{itemize}

The rest of this paper is organized as follows. Section II details our system model and adopted assumptions. In Section III, we examine the performance the truncated CIPC scheme in the context of covert communications. Section IV presents our analysis on the conventional CIPC scheme. Section V provides numerical results to confirm our analysis and provide useful insights with regard to the comparison between these two schemes. Finally, conclusions are drawn in Section VI.

\emph{Notation:} Scalar variables are denoted by italic symbols. Vectors are denoted by lower-case boldface symbols. Given a complex vector, $(\cdot)^{\dag}$ denotes the conjugate transpose. Given a complex number, $|\cdot|$ denotes its modulus and $(\cdot)^{\ast}$ denotes the conjugation.

\section{System Model}
\subsection{Considered Communication Scenario}
\begin{figure} [ht!]
  \centering
  \includegraphics[scale=0.8]{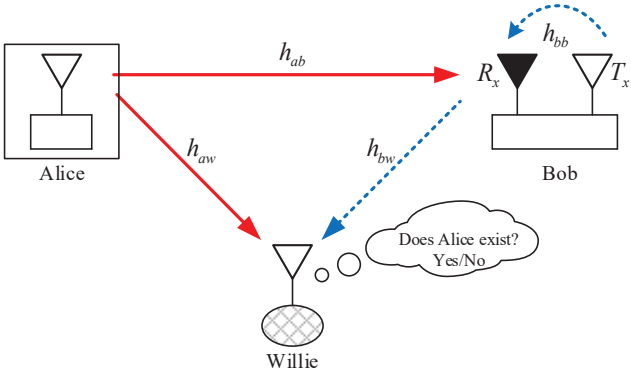}\\
  \caption{Covert communications network model.}\label{fig1}
\end{figure}

As shown in Fig.~\ref{fig1}, we consider a wireless communication scenario, where Alice (i.e., the transmitter) wants to transmit information covertly to Bob, while trying to hide herself from Willie (i.e., the warden). Meanwhile, Willie is to detect Alice's transmission by observing the wireless environment. We assume the wireless channels within our system model are subject to independent quasi-static Rayleigh fading with equal block length, which means that all the channel coefficients are independent and identically distributed (i.i.d.) circularly symmetric complex Gaussian random variables with zero-mean and unit-variance. Time is slotted and the quasi-static Rayleigh fading channel coefficient remains fixed for a slot. Alice and Willie are assumed to have a single antenna, while besides the single receiving antenna, Bob uses an additional antenna for transmission of
AN in order to deliberately confuse Willie. The channel from Bob to Alice, Bob to Willie, Bob to Bob, Alice to Bob, or Alice to Willie, is denoted by $h_j$ and the mean value of $|h_j|^2$ is denoted by $\lambda_j$, where the subscript $j$ can be $ba, bw, bb, ab, aw$, respectively, corresponding to different channels. In this work, from a conservative point of view we assume that $\lambda_j$ is publicly known, since, for example, Willie can possibly know the location of Bob and the potential area or location (e.g., a building) of Alice.

In order to achieve the ultimate goal of hiding the presence of Alice, in the considered scenario as a base station Bob broadcasts pilots periodically in order to enable Alice's estimation of the channel from Bob to Alice (i.e., $h_{ba}$). Meanwhile, Willie can estimate the channel from Bob to Willie (i.e., $h_{bw}$), since the pilots transmitted by Bob are publicly known. Considering channel reciprocity, we assume the channel from Alice to Bob is the same as $h_{ba}$. Since Alice does not transmit any pilots, it is assumed that Willie does not known $h_{aw}$. To eliminate the requirement that Bob has to know $h_{ab}$ for decoding and avoid the feedback from Alice to Bob, in this work we consider the CIPC at Alice, in which Alice varies its transmit power as per $h_{ab}$ such that $P_a|h_{ab}|^2$ is a fixed value, i.e.,
\begin{align}\label{Q_definition}
P_a|h_{ab}|^2 = Q,
\end{align}
where $P_a$ is the transmit power of Alice. Following \cite{Hu2017ICC}, we note that our current analysis in this work is significantly different from \cite{Hu2017ICC}, although the considered scenario is similar. This is due to the following two facts.  The first fact is that in \cite{Hu2017ICC} it is assumed that Willie knows the channel from Alice to Willie (i.e., $h_{aw}$), while in this section we assume that $h_{aw}$ is unknown to Willie (since Alice did not transmit any signal before the covert transmission). The second fact is that in this work we adopt the CIPC at Alice, while in \cite{Hu2017ICC} a fixed power is adopted at Alice.

It should be noted that the proposed CIPC scheme is to achieve covert communications in wireless fading channels without the need for Alice to transmit training signals for Bob to perform channel estimation. This can ensure the existence of Alice is hidden from being detected. Meanwhile, using the constant transmit power scheme commonly requires Alice to transmit training signals for Bob to estimate the channel, which easily reveals Alice' s existence. Therefore it is not appropriate to include constant transmit power scheme for comparison.

Considering some specific quality of service (QoS) requirements, we consider that the transmission throughput from Alice to Bob is fixed and predetermined, which is denoted by $R$. As such, transmission from Alice to Bob incurs outage when $C_{ab}<R$, where $C_{ab}$ is the channel capacity from Alice to Bob. Then, the transmission outage probability is given by
\begin{align}\label{outage_definition}
\delta = \mathcal{P}[C_{ab}<R].
\end{align}
As such, in this work we adopt the effective transmission throughput as the main performance metric for the communication from Alice to Bob.

Next, we detail the transmission from Alice to the full-duplex Bob and derive the associated transmission outage probability.
As a practical technique, the full-duplex radio has been widely explored in wireless communications (e.g., \cite{Duarte2012Experiment,Sabharwal2014In-band}). For example, using a full-duplex receiver to generate AN for security enhancement has been widely examined  the context of physical layer security (e.g., \cite{Zheng2013Improving,Yan2018Secretpilot,Hu2017ICC}).
Similarly to \cite{Hu2017ICC}, we consider a full-duplex receiver Bob who simultaneously receives covert information signal from Alice and transmits AN with a random power to confuse the warden Willie on detecting the covert transmission. Generating AN by the full-duplex Bob enables varying the presence and amount of uncertainty at Willie on demand, which leads to that the covert communication between Alice and Bob is fully under the control of Alice and Bob.
When Alice transmits the covert information, the signal received at the full-duplex Bob is given by
\begin{align} \label{y_b}
\mathbf{y}_b[i]=\sqrt{P_a}h_{ab}\mathbf{x}_a[i]+\sqrt{\phi P_b}h_{bb}\mathbf{v}_b[i]+\mathbf{n}_b[i],
\end{align}
where $\mathbf{x}_a$ is the signal transmitted by Alice, and $\mathbf{n}_b[i]$ is the AWGN at Bob with $\sigma^2_b$ as its variance, i.e., $\mathbf{n}_b[i] \thicksim\mathcal{CN}(0,\sigma^2_b)$, $\mathbf{v}_b$ is the AN signal transmitted by Bob satisfying $\mathbb{E}[\mathbf{v}_b[i]\mathbf{v}^{\dag}_b[i]]=1$, $h_{bb}$ denotes the self-interference channel at Bob (i.e., the channel between Bob's transmitting antenna and Bob's receiving antenna), and $P_b$ is Bob's transmit power of the AN signal.
In this work, we adopt a generalized self-interference cancellation model, in which the self-interference can be fully cancelled (i.e., $\phi = 0$) or cannot be fully cancelled (i.e., $0 < \phi\leq1$), where $\phi$ in \eqref{y_b} denotes the self-interference cancellation coefficient \cite{Elsayed2015All}. In practice, the self-interference may not be fully cancelled due to the fact that self-interference channel may not be perfectly estimated in some specific scenarios \cite{Kim2013Effects}. We note that $0\leq\phi\leq1$ corresponds to different self-interference cancellation levels~\cite{Everett2014Passive}.
In order to remove the impact of the phase in the complex channel coefficient, Alice has to pre-process the transmitted signal $\mathbf{x}_a$ as
\begin{equation}
\mathbf{x}_a=\frac{h_{ab}^{\ast}}{|h_{ab}|}\mathbf{x}_{r},
\end{equation}
where $\mathbf{x}_{r}$ is the raw information signal satisfying $\mathbb{E}[\mathbf{x}_a[i]\mathbf{x}^{\dag}_a[i]]=1$, $i = 1, 2, \dots, n$ is the index of each channel use.

In order to create uncertainty at Willie, in this work we assume that $P_b$ changes randomly from slot to slot \cite{Sobers2015Covert,Sobers2017Covert} and and follows a continuous uniform distribution over the interval $[0, P_b^{\max}]$ with pdf given by
\begin{equation}\label{pb_pdf}
\begin{aligned}
f_{P_b}(x)=
\begin{cases}
\frac{1}{P_b^{\mathrm{max}}} & \text{if}\quad 0\leq x\leq P_b^{\mathrm{max}}, \\
0, & \text{otherwise}.
\end{cases}
\end{aligned}
\end{equation}
Since Willie possesses the knowledge of $h_{bw}$ in the slot under consideration, for a constant transmit power at Bob, it is straightforward for him to detect the covert transmission when the additional power (on top of his receiver noise power) is received. The purpose of introducing randomness in Bob's transmit power is to create uncertainty in Willie's received power, such that Willie is unsure whether an increase in his received power is due to Alice's transmission or simply due to a variation in Bob's transmit power of the AN signal. We note that we consider the uniform distribution as an initial example and other distributions will be explored in our future works.

\subsection{Detection Performance at Willie}

In the considered communication scenario, Willie is to infer the presence of Alice by detecting the wireless transmission from Alice to Bob. As such, Willie has a binary detection problem, in which Alice does not transmit information to Bob in the null hypothesis $\Hnull$ but it does in the alternative hypothesis $\Halt$. The ultimate goal for Willie is to detect whether his observation comes from $\Hnull$ or $\Halt$ by applying some specific decision rule. The detection performance of Willie is normally measured by the detection error probability, which is defined as
\begin{align} \label{P_e}
\mathcal{P}_e\triangleq \pi_0\alpha + \pi_1\beta,
\end{align}
where $\mathcal{P}(\Halt)=\pi_0$ is the probability that Alice transmits a covert message, $\mathcal{P}(\Hnull)=\pi_1$ is the probability that Alice does not transmit a covert message, $\alpha=\mathcal{P}(\Hoalt|\Hnull)$ is the false alarm rate and $\beta=\mathcal{P}(\Honull|\Halt)$ is the miss detection rate, while $\Hoalt$ and $\Honull$ are the binary decisions that infer whether Alice transmits information to Bob or not, respectively.

From a conservative point of view, in this work we consider the worst-case scenario for Alice and Bob, where Willie has the ability to develop the detection strategy and the optimal detection threshold on it. As per \cite{Boulat2016Time,Sobers2017Covert}, $\mathcal{P}_e \geq \min\{\pi_0, \pi_1\}(\alpha+\beta)$. As such, Alice can achieve covert communication, for any $\epsilon > 0$, $\min\{\pi_0, \pi_1\}\xi \geq \min\{\pi_0, \pi_1\}-\epsilon$  for $n$ sufficiently large, where $\xi$ is the total error probability, which is given by
\begin{align} \label{xi_definition}
\xi=\alpha + \beta,
\end{align}
and $\epsilon \in [0,1]$ is a predetermined value to specify the covert communication constraint.


Next, we derive the false alarm and miss detection rates at Willie, based on which we derive the optimal detection threshold that minimizes the total error probability. We focus on one communication slot, where Willie has to decide whether Alice transmitted to Bob, or not. Thus Willie faces a binary hypothesis testing problem. The composite received signal model at Willie is given by
\begin{eqnarray}\label{yw_FD}
\mathbf{y}_w[i]\!=\!
 \left\{ \begin{aligned}
        \ &\sqrt{P_b}h_{bw}\mathbf{v}_b[i]+\mathbf{n}_w[i], ~~~~~~~~~~~~~~~~~~~~~\Hnull, \\
        \ &\sqrt{P_a}h_{aw}\mathbf{x}_a[i]+\sqrt{P_b}h_{bw}\mathbf{v}_b[i]+\mathbf{n}_w[i],  ~~\Halt.
         \end{aligned} \right.
\end{eqnarray}
In this work we assume that Willie employs a radiometer as his detection test \cite{lee2015achieving}. Considering the infinite blocklength, i.e., $n \rightarrow \infty$, we have
\begin{eqnarray}\label{t_w_FD}
T \!\triangleq \!\lim_{n \rightarrow \infty} \frac{1}{n}T(n)\!=\!
 \left\{ \begin{aligned}
        \ &P_b|h_{bw}|^2+\sigma_w^2, ~~~~~~~~~~~~~~~~\Hnull, \\
        \ &P_a|h_{aw}|^2+P_b|h_{bw}|^2+\sigma_w^2, ~~\Halt.
         \end{aligned} \right.
\end{eqnarray}
Then, the decision rule embedded in the detector at Willie is given by
\begin{align}\label{decisions_FD}
T\mathop{\gtrless}\limits_{\Honull}^{\Hoalt}\tau,
\end{align}
where $\tau$ is the detection threshold for $T$, which will be optimally determined in order to minimize the total error probability.

\section{Truncated Channel Inversion Power Control Scheme}

In this section, we examine the possibility and performance of covert communications by utilizing AN generated by the full-duplex Bob (i.e., the receiver), since this AN can lead to a certain amount of uncertainty at Willie. Generating AN by the full-duplex Bob enables varying the presence and amount of uncertainty at Willie on demand, which leads to that the covert communication between Alice and Bob is fully under the control of Alice and Bob.

Considering a practical scenario, the truncated CIPC scheme is considered at Alice in this section, where Alice can only transmit covert information when the quality of the channel from Alice to Bob (i.e., $|h_{ab}|^2$) is greater than some specific value \cite{Hesham2014Truncated}. As such, Alice's transmit power is given by
\begin{align} \label{truncated inversion policy}
P_a=\left\{
  \begin{array}{ll}
    \frac{Q}{|h_{ab}|^2}, & |h_{ab}|^2\geq \frac{Q}{P_a^{\mathrm{max}}}, \\
    0,  &|h_{ab}|^2< \frac{Q}{P_a^{\mathrm{max}}},
  \end{array}
\right.
\end{align}
where $P_a^{\mathrm{max}}$ is the maximum power constraint at Alice. As per \eqref{truncated inversion policy}, we note that Alice can transmit the covert message when $|h_{ab}|^2 \geq Q/P_a^{\mathrm{max}}$ is met. We denote this necessary condition as $\mathbb{C}$. As such, Alice can transmit $\mathbf{x}_a$ to Bob whenever condition $\mathbb{C}$ is met. Considering quasi-static Rayleigh fading, the cumulative distribution function (CDF) of $|h_{ab}|^2$ is given by $F_{|h_{ab}|^2}(x) = 1 - e^{-x/\lambda_{ab}}$ and thus the probability that $\mathbb{C}$ is guaranteed is given by
\begin{align} \label{condition_c}
\mathcal{P}_\mathbb{C} = \exp\left\{-\frac{Q}{\lambda_{ab}P_a^{\mathrm{max}}}\right\}.
\end{align}

\subsection{Detection Performance at Willie}

As discussed before, Alice can transmit covert message when condition $\mathbb{C}$ is guaranteed. In the truncated CIPC scheme, we assume that Alice will transmit a covert message with probability 1/2 when $\mathbb{C}$ is true. As per \eqref{condition_c}, the probabilities $\mathcal{P}(\Halt)$ and $\mathcal{P}(\Hnull)$ are, respectively, given by
\begin{align} \label{PH1_trancated}
\mathcal{P}(\Halt)&=\pi_1=\frac{1}{2}\mathcal{P}_\mathbb{C} \notag \\
&=  \frac{1}{2}\exp\left\{-\frac{Q}{\lambda_{ab}P_a^{\mathrm{max}}}\right\},
\end{align}
and
\begin{align} \label{PH0_trancated}
\mathcal{P}(\Hnull)&=\pi_0  \notag \\
&= 1-\frac{1}{2}\exp\left\{-\frac{Q}{\lambda_{ab}P_a^{\mathrm{max}}}\right\}.
\end{align}

Following the decision rule given in \eqref{decisions_FD}, we derive the false alarm and miss detection rates at Willie in the following theorem.
\begin{theorem}\label{theorem2_trancated}
The false alarm and miss detection rates at Willie are derived in \eqref{PFA_FD_trancated} and \eqref{PMD_FD_trancated}, respectively.
\begin{align}
\alpha&=\left\{
  \begin{array}{ll}
    1,  &\tau<\sigma_w^2, \\
    1-\frac{\tau-\sigma_w^2}{P_b^{\mathrm{max}}|h_{bw}|^2}, & \sigma_w^2\leq\tau\leq\nu,\\
    0,  &\tau>\nu,
  \end{array}
\right. \label{PFA_FD_trancated}
\end{align}

\setcounter{equation}{16}
where
\begin{align}\label{nu_definition}
&\nu\triangleq P_b^{\mathrm{max}}|h_{bw}|^2+\sigma_w^2,
\end{align}
and the exponential integral function $\mathrm{Ei}(\cdot)$ is given by
\begin{align} \label{Ei_definition}
\mathrm{Ei}(x)=-\int_{-x}^{\infty} \frac{e^{-t}}{t} \mathrm{d}t.
\end{align}
\end{theorem}

\begin{IEEEproof}
Following \eqref{t_w_FD} and \eqref{decisions_FD}, and noting that $\Hnull$ can happen regardless of whether the condition $\mathbb{C}$ is guaranteed, the false alarm rate is given by
\begin{align}
\alpha&=\mathcal{P}\left[P_b|h_{bw}|^2+\sigma_w^2>\tau\big{|}\mathbb{C}\right]+\mathcal{P}\left[P_b|h_{bw}|^2+\sigma_w^2>\tau\big{|}\mathbb{C}^{\prime}\right] \notag \\
&=\mathcal{P}\left[P_b|h_{bw}|^2+\sigma_w^2>\tau\right]  \notag \\
&=\left\{
  \begin{array}{ll}
    1,  &\tau<\sigma_w^2, \\
    \int_{\frac{\tau-\sigma_w^2}{|h_{bw}|^2}}^{P_b^{\mathrm{max}}}f_{P_b}(x)\mathrm{d}x, &\sigma_w^2\leq\tau\leq\nu,\\
    0,  &\tau>\nu,
  \end{array}
\right.\notag \\
&=\left\{
  \begin{array}{ll}
    1,  &\tau<\sigma_w^2, \\
    \int_{\frac{\tau-\sigma_w^2}{|h_{bw}|^2}}^{P_b^{\mathrm{max}}}\frac{1}{P_b^{\mathrm{max}}}\mathrm{d}x, &\sigma_w^2\leq\tau\leq\nu,\\
    0,  &\tau>\nu.
  \end{array}
\right. \label{fa_proof2_trancated}
\end{align}
Then, solving the integral in \eqref{fa_proof2_trancated} leads to the desired result in \eqref{PFA_FD_trancated}.

As per \eqref{t_w_FD} and \eqref{decisions_FD}, the miss detection rate is given by
\begin{align}
\beta&=\mathcal{P}\left[P_a|h_{aw}|^2+P_b|h_{bw}|^2+\sigma_w^2<\tau\big{|}\mathbb{C}\right] \notag \\
&=\mathcal{P}\left[\frac{Q|h_{aw}|^2}{|h_{ab}|^2}+P_b|h_{bw}|^2+\sigma_w^2<\tau\big{|}\mathbb{C}\right] \notag \\
&=\left\{
\begin{array}{ll}
    0,  &\tau<\sigma_w^2, \\
    \mathcal{P}_\mathbb{C}^{-1}\mathcal{P}\left[P_b<\frac{\tau-\sigma_w^2-{Q|h_{aw}|^2}/{|h_{ab}|^2}}{|h_{bw}|^2}\right], &\sigma_w^2\leq\tau\leq\nu,\\
    \mathcal{P}_\mathbb{C}^{-1}\mathcal{P}\left[\frac{|h_{aw}|^2}{|h_{ab}|^2}<\frac{\tau-\sigma_w^2-P_b|h_{bw}|^2}{Q}\right],  &\tau>\nu,
\end{array}
\right.  \label{md_proof3_trancated}
\end{align}
Following \eqref{md_proof3_trancated}, we achieve the desired result in \eqref{PMD_FD_trancated} after some algebra manipulations.
\end{IEEEproof}

We note that the false alarm and miss detection rates derived in Theorem~\ref{theorem2_trancated} are for an arbitrary detection threshold $\tau$. In practice, Willie will determine the optimal detection threshold that minimizes the detection error probability given in \eqref{P_e}. Considering that the probabilities $\mathcal{P}(\Halt)$ and $\mathcal{P}(\Hnull)$ are not functions of the detection threshold at Willie, the optimal detection threshold is also the one that minimizes the total error probability (i.e., $\alpha + \beta$) given by \eqref{xi_definition}, which is derived in the following theorem.

\begin{proposition}\label{proposition4_trancated}
For the decision rule given in \eqref{decisions_FD}, Willie's optimal threshold that minimizes the total error probability is derived as
\begin{align}
\tau^{\ast}=\nu,
\end{align}
and the corresponding minimum total error probability is given by
\begin{align} \label{xi ast_trancated}
\xi^{\ast}&=1\!-\!\frac{Q\lambda_{aw}\exp\left(\frac{Q}{P_a^{\max}\lambda_{ab}}\right)}{P_b^{\max}\lambda_{ab}|h_{bw}|^2}\times \\
&~~~\left[\mathrm{Ei}\left(-\frac{P_b^{\max}|h_{bw}|^2\lambda_{ab}+Q\lambda_{aw}}{P_a^{\max}\lambda_{aw}\lambda_{ab}}\right)\!-\!\mathrm{Ei}\left(-\frac{Q}{P_a^{\max}\lambda_{ab}}\right)\right].\notag
\end{align}
\end{proposition}
where $\nu = P_b^{\mathrm{max}}|h_{bw}|^2+\sigma_w^2$ as defined in \eqref{nu_definition}.

\newcounter{mytempeqncnt2}
\begin{figure*}[tp]
\normalsize
\setcounter{mytempeqncnt2}{\value{equation}}
\setcounter{equation}{15}
\begin{align}
\beta&=\left\{
\begin{array}{ll}
    0,  &\tau<\sigma_w^2, \\
    \frac{\tau-\sigma_w^2}{P_b^{\max}|h_{bw}|^2}-\frac{Q\lambda_{aw}\exp\left(\frac{Q}{P_a^{\max}\lambda_{ab}}\right)}{P_b^{\max}\lambda_{ab}|h_{bw}|^2}\left[\mathrm{Ei}\left(-\frac{(\tau-\sigma_w^2)\lambda_{ab}+Q\lambda_{aw}}{P_a^{\max}\lambda_{aw}\lambda_{ab}}\right)-\mathrm{Ei}\left(-\frac{Q}{P_a^{\max}\lambda_{ab}}\right)\right], &\sigma_w^2\leq\tau\leq\nu,\\
    1-\frac{Q\lambda_{aw}\exp\left(\frac{Q}{P_a^{\max}\lambda_{ab}}\right)}{P_b^{\max}|h_{bw}|^2\lambda_{ab}}\Big[\mathrm{Ei}\left(-\frac{(\tau-\sigma_w^2)\lambda_{ab}+Q\lambda_{aw}}{P_a^{\max}\lambda_{aw}\lambda_{ab}}\right)-\mathrm{Ei}\left(-\frac{(\tau-\nu)\lambda_{ab}+Q\lambda_{aw}}{P_a^{\max}\lambda_{aw}\lambda_{ab}}\right)\Big],  &\tau>\nu,
\end{array}
\right.  \label{PMD_FD_trancated}
\end{align}
\setcounter{equation}{22}
\begin{align} \label{xi_case_trancated}
\xi&=\left\{
\begin{array}{ll}
    1,  &\tau<\sigma_w^2, \\
    1-\frac{Q\lambda_{aw}\exp\left(\frac{Q}{P_a^{\max}\lambda_{ab}}\right)}{P_b^{\max}\lambda_{ab}|h_{bw}|^2}\left[\mathrm{Ei}\left(-\frac{(\tau-\sigma_w^2)\lambda_{ab}+Q\lambda_{aw}}{P_a^{\max}\lambda_{aw}\lambda_{ab}}\right)-\mathrm{Ei}\left(-\frac{Q}{P_a^{\max}\lambda_{ab}}\right)\right],
     &\sigma_w^2\leq\tau\leq\nu,\\
    1-\frac{Q\lambda_{aw}\exp\left(\frac{Q}{P_a^{\max}\lambda_{ab}}\right)}{P_b^{\max}\lambda_{ab}|h_{bw}|^2}\left[\mathrm{Ei}\left(-\frac{(\tau-\sigma_w^2)\lambda_{ab}+Q\lambda_{aw}}{P_a^{\max}\lambda_{aw}\lambda_{ab}}\right)-\mathrm{Ei}\left(-\frac{(\tau-\nu)\lambda_{ab}+Q\lambda_{aw}}{P_a^{\max}\lambda_{aw}\lambda_{ab}}\right)\right], &\tau>\nu.
\end{array}
\right.
\end{align}
\setcounter{equation}{\value{mytempeqncnt2}}
\hrulefill
\vspace*{4pt}
\end{figure*}

\setcounter{equation}{23}

\begin{IEEEproof}
As per \eqref{xi_definition}, \eqref{PFA_FD_trancated}, and \eqref{PMD_FD_trancated}, the total error probability at Willie is given by \eqref{xi_case_trancated}.

We first note that $\xi = 1$ is the worst case scenario for Willie and thus Willie does not set $\tau < \sigma_w^2$.
As per \eqref{xi ast_trancated}, for $\sigma_w^2 \leq \tau\leq\nu$ the total error probability $\xi$ is a monotonically decreasing function of $\tau$. Thus, Willie will set $\nu$ as the threshold to minimize $\xi$ in this case. For $\tau>\nu$, we have
\begin{align} \label{first_derivative_trancated}
\frac{\partial \xi}{\partial \tau}&=\frac{Q\lambda_{aw}\exp\left(\frac{Q}{P_a^{\max}\lambda_{ab}}\right)}{P_b^{\max}|h_{bw}|^2\lambda_{ab}}\left[g(-\kappa_1)-g(-\kappa_2)\right],
\end{align}
where
\begin{align}\label{g_definition}
g(x)&\triangleq\frac{e^{x}}{x},  \\
\kappa_1&\triangleq\frac{(\tau-\sigma_w^2)\lambda_{ab}+Q\lambda_{aw}}{P_a^{\max}\lambda_{aw}\lambda_{ab}}, \notag\\
\kappa_2&\triangleq\frac{(\tau-\nu)\lambda_{ab}+Q\lambda_{aw}}{P_a^{\max}\lambda_{aw}\lambda_{ab}}. \notag
\end{align}
In order to check the sign of $g(-\kappa_1)-g(-\kappa_2)$ in \eqref{first_derivative_trancated}, we derive the first derivative of $g(x)$ with respect to $x$ as
\begin{align} \label{gx_first derivative}
\frac{\partial g(x)}{\partial x}&=\frac{e^{x}(x-1)}{x^2}.
\end{align}
As per \eqref{gx_first derivative} we can see that ${\partial g(x)}/{\partial x}<0$ for $x<0$. Noting that $\kappa_1>\kappa_2$, we have $g(-\kappa_1)-g(-\kappa_2) >0$. Noting that $\kappa_1>\kappa_2$, the value of the term $g(-\kappa_1)-g(-\kappa_2)$ is larger than 0, which leads to the value of ${\partial \xi}/{\partial \tau}>0$. As such, $\xi$ is a monotonically increasing function of $\tau$ based on \eqref{xi_case_trancated}, which indicates that Willie will try to set $\nu$ as the threshold again to minimize $\xi$ under this case. We also note that for the two cases, i.e., $\sigma_w^2 \leq \tau\leq\nu$ and $\tau>\nu$, setting $\tau = \nu$ can achieve the same total error probability. As such, we can conclude that the optimal detection threshold is $\tau$.
\end{IEEEproof}

Following Theorem~\ref{theorem2_trancated} and Proposition~\ref{proposition4_trancated}, we note that although the noise power at Willie i.e., $\sigma_w^2$ appears in the expressions of the false alarm rate, miss detection rate, and the optimal detection threshold, it does not affect the minimum total error probability, $\xi^{\ast}$, at Willie. This is due to the fact that Willie knows $\sigma_w^2$ and thus he can adjust the optimal detection threshold accordingly to counteract the impact of $\sigma_w^2$. We also note that for $\tau^{\ast} = \nu$ the false alarm rate is zero, which means that Willie adjusts the detection threshold to force the false alarm rate being zero in order to minimize the total error probability. This is achievable, since as assumed Willie knows the value range of Bob's transmit power of the AN signal (i.e., the maximum value of $P_b$, which is $P_b^{\mathrm{max}}$).

\subsection{Transmission from Alice to Bob}

In this subsection, we detail the transmission from Alice to the full-duplex Bob and derive the
associated transmission outage probability.

Following~\eqref{y_b}, the signal to interference plus noise ratio (SINR) at Bob used to decode $\mathbf{x}_a$ is given by
\begin{align} \label{gamma_b}
\gamma_b=\frac{P_a|h_{ab}|^2}{\phi P_b |h_{bb}|^2+\sigma_b^2} =\frac{Q}{\phi P_b |h_{bb}|^2+\sigma_b^2},
\end{align}
since Alice varies its transmit power as per $h_{ab}$ such that $P_a|h_{ab}|^2 = Q$. Due to the randomness in $|h_{bb}|^2$ and $P_b$, we note that the residual error after self-interference cancellation is complex Gaussian and its variance is proportional to $P_b |h_{bb}|^2$. Then, the transmission from Alice to Bob will incur outage when $C_{ab}<R$, where $C_{ab} = \log_2(1 + \gamma_b)$ is the channel capacity from Alice to Bob and $R$ is the fixed transmission throughput from Alice to Bob. We note that $\gamma_b$ cannot be larger than ${Q}/{\sigma_b^2}$, which is achieved when $\phi=0$ (i.e., when the self-interference can be fully cancelled). As such, in order to guarantee the transmission outage probability being less than one, we to have to ensure
$R < \log_2\left(1+{Q}/{\sigma_b^2} \right)$, which is assumed to be true in this section.
We next derive the transmission outage probability from Alice to Bob in the following proposition.

\begin{proposition}\label{proposition3} The transmission outage probability from Alice to the full-duplex Bob is derived as
\begin{align} \label{outage_delta}
\delta = e^{-\eta}+\eta \mathrm{Ei}(-\eta),
\end{align}
where
\begin{align}\label{eta_definition}
\eta=\frac{Q-(2^{R}-1)\sigma_b^2}{\left(2^{R}-1\right)\lambda_{bb}\phi P_b^{\mathrm{max}}}.
\end{align}
\end{proposition}
\begin{IEEEproof} Based on the definition of the transmission outage probability given in \eqref{outage_definition}, we have
\begin{align} \label{calulate_delta}
\delta&=\mathcal{P}\left\{\log_2\left(1+\gamma_b\right)\leq R \right\} \notag \\
&=\mathcal{P}\left\{\frac{Q}{\phi P_b |h_{bb}|^2+\sigma_b^2}\leq 2^{R}-1\right\} \notag \\
&=\int_0^{P_b^{\mathrm{max}}}\int_{\frac{Q-(2^R-1)\sigma_b^2}{(2^R-1)\phi y}}^{+\infty}f_{|h_{bb}|^2}(x)f_{P_b}(y)\mathrm{d}x \mathrm{d}y \notag \\
&=\int_0^{P_b^{\mathrm{max}}}\int_{\frac{Q-(2^R-1)\sigma_b^2}{(2^R-1)\phi y}}^{+\infty}\frac{\exp\left(-\frac{x}{\lambda_{bb}}\right)}{\lambda_{bb}P_b^{\mathrm{max}}}\mathrm{d}x \mathrm{d}y \notag \\
&=\frac{1}{P_b^{\mathrm{max}}}\int_0^{P_b^{\mathrm{max}}}\exp\left\{-\frac{Q-(2^{R}-1)\sigma_b^2}{(2^{R}-1)\lambda_{bb}\phi y}\right\}\mathrm{d}y \notag \\
&=\exp\left(-\frac{Q-(2^{R}-1)\sigma_b^2}{(2^{R}-1)\lambda_{ab}\phi P_b^{\mathrm{max}}}\right)+\frac{Q-(2^{R}-1)\sigma_b^2}{(2^{R}-1)\lambda_{ab}\phi P_b^{\mathrm{max}}}\times  \notag \\
&~~~\mathrm{Ei}\left(-\frac{Q-(2^{R}-1)\sigma_b^2}{(2^{R}-1)\lambda_{ab}\phi P_b^{\mathrm{max}}}\right).
\end{align}
Following \eqref{calulate_delta}, we achieve the desired result in \eqref{outage_delta} after some algebra manipulations.
\end{IEEEproof}

Following Proposition~\ref{proposition3}, we determine some properties of the transmission outage probability in the following corollary.

\begin{corollary}\label{corollary1}
The transmission outage probability $\delta$ is a monotonically decreasing function of $\eta$, which leads to the fact that $\delta$ monotonically decreases with $Q$ but increases with $P_b^{\mathrm{max}}$.
\end{corollary}

\begin{IEEEproof}
In order to determine the monotonicity of $\delta$ with respect to $\eta$, we derive its first derivative as
\begin{align}
\frac{\partial \delta}{\partial \eta}&=-2e^{-\eta}+\mathrm{Ei}(-\eta).
\end{align}
We note that $\eta \geq 0$ as per its definition given in \eqref{eta_definition} and thus we have ${\partial \delta}/{\partial \eta} \leq 0$ due to $-2e^{-\eta}<0$ and $\mathrm{Ei}(-\eta)<0$, which indicates that $\delta$ monotonically decreases with $\eta$. Again, as per the definition of $\eta$ in \eqref{eta_definition}, we can see that $\eta$ is a monotonically increasing function of $Q$ but a monotonically decreasing function of $P_b^{\mathrm{max}}$, which completes the proof of Corollary~\ref{corollary1}.
\end{IEEEproof}

\subsection{Optimization of Effective Covert Throughput}

In this subsection, we examine the maximum ECT achieved in the considered scenario with the full-duplex Bob subject to a certain covert communication constraint. In this scenario, Willie knows $h_{bw}$, which is the reason that the detection performance (e.g., false alarm rate, miss detection rate, optimal detection threshold, and the minimum total error probability) at Willie depends on $h_{bw}$. However, Alice does not know $h_{bw}$ and thus cannot guarantee the covert communication constraint $\min \{\pi_0, \pi_1\}\xi^{\ast}(|h_{aw}|^2,Q) \geq \min \{\pi_0, \pi_1\} - \epsilon$. As such, in the following proposition we present the expected value of $\xi^{\ast}(|h_{aw}|^2,Q)$ over all realizations of $|h_{bw}|^2$, which is denoted by $\overline{\xi^{\ast}}(Q)$, and then we use $\min \{\pi_0, \pi_1\}\overline{\xi^{\ast}}(Q) \geq \min \{\pi_0, \pi_1\} - \epsilon$ as the covert communication constraint in this section.

%

The optimization problem at Alice of maximizing the ECT subject to a certain covert communication constraint is given by
\begin{align}\label{P1_trancated}
Q^{\ast} = &\argmax_{Q} R_c \\ \nonumber
&\text{s. t.} \quad  \min\{\pi_0, \pi_1\}\overline{\xi^{\ast}}(Q) \geq \min\{\pi_0, \pi_1\}- \epsilon,
\end{align}
where $R_c$ is the effective rate from Alice to Bob (without consider the factor of 1/2), which is given by
\begin{align}
R_c = R(1-\delta)\mathcal{P}_\mathbb{C},
\end{align}
and $\overline{\xi^{\ast}}(Q)$ is the expected value of the minimum total error probability $\xi^{\ast}(|h_{bw}|^2,Q)$ over all realizations of $|h_{bw}|^2$, which is given by
\begin{align}\label{expected_xi_proof_trancated}
&\overline{\xi^{\ast}}(Q)=\int_0^{\infty} \xi^{\ast}(|h_{bw}|^2,Q)\frac{e^{-\frac{|h_{bw}|^2}{\lambda_{bw}}}}{\lambda_{bw}} \mathrm{d} |h_{bw}|^2  \\
&=1\!-\!\int_0^{\infty}\!\frac{Q\lambda_{aw}\!\exp\!\left(\frac{Q}{P_a^{\max}\lambda_{ab}}\!-\!\frac{|h_{bw}|^2}{\lambda_{bw}}\right)}
{P_b^{\max}\lambda_{ab}\lambda_{bw}|h_{bw}|^2}\!\times \notag \\
&\left[\mathrm{Ei}\!\left(\!-\!\frac{P_b^{\max}|h_{bw}|^2\lambda_{ab}\!+\!Q\lambda_{aw}}{P_a^{\max}\lambda_{aw}\lambda_{ab}}\right)\!-\!\mathrm{Ei}\!\left(\!-\!\frac{Q}{P_a^{\max}\lambda_{ab}}\right)\right]\!\mathrm{d} |h_{bw}|^2. \notag
\end{align}
The objective function  $R_c=R(1-\delta)\mathcal{P}_\mathbb{C}$ in \eqref{P1_trancated} is not a monotonic function of $Q$. This is because that the outage $\delta$ is a monotonically decreasing function of $Q$ as per Corollary~\ref{corollary1} and $\mathcal{P}_\mathbb{C}$ is a monotonically decreasing function of $Q$ as per\eqref{condition_c}. The optimization problem \eqref{P1_trancated} can be solved by numerical search in the set of Q which satisfies $1/2\exp(-Q/(\lambda_{ab}P_a^{\mathrm{max}}))[1-\overline{\xi^{\ast}}(Q)] \leq \epsilon$.

\newcounter{mytempeqncnt3}
\begin{figure*}[tp]
\normalsize
\setcounter{mytempeqncnt2}{\value{equation}}
\setcounter{equation}{37}
\begin{align}
\beta&=\left\{
  \begin{array}{ll}
    0,  &\tau<\sigma_w^2, \\
    \frac{1}{P_b^{\mathrm{max}}|h_{bw}|^2}\left(\tau-\sigma_w^2-\frac{Q\lambda_{aw}}{\lambda_{ab}}\ln\left(1+\frac{(\tau-\sigma_w^2)\lambda_{ab}}{Q\lambda_{aw}}\right)\right), & \sigma_w^2\leq\tau\leq\nu,\\
    1-\frac{Q\lambda_{aw}}{P_b^{\mathrm{max}}|h_{bw}|^2\lambda_{ab}}\ln\left(1+\frac{P_b^{\mathrm{max}}|h_{bw}|^2}{\tau+Q\lambda_{aw}/\lambda_{ab}-\nu}\right),  &\tau>\nu,
  \end{array}
\right. \label{PMD_FD}
\end{align}
\setcounter{equation}{42}
\begin{align} \label{xi_case}
\xi=\left\{
  \begin{array}{ll}
    1, & \tau < \sigma_w^2, \\
    1-\frac{Q\lambda_{aw}}{P_b^{\mathrm{max}}|h_{bw}|^2\lambda_{ab}}\ln\left(1+\frac{(\tau-\sigma_w^2)\lambda_{ab}}{Q\lambda_{aw}}\right),  &\sigma_w^2 \leq\tau\leq\nu, \\
    1-\frac{Q\lambda_{aw}}{P_b^{\mathrm{max}}|h_{bw}|^2\lambda_{ab}}\ln\left(1+\frac{P_b^{\mathrm{max}}|h_{bw}|^2}{\tau+Q\lambda_{aw}/\lambda_{ab}-\nu}\right),  &\tau>\nu.
  \end{array}
\right.
\end{align}
\setcounter{equation}{\value{mytempeqncnt3}}
\hrulefill
\vspace*{4pt}
\end{figure*}

\section{Conventional Channel Inversion Power Control Scheme}

In this section, we consider the conventional CIPC scheme in the context of covert communications. The conventional CIPC scheme is a special case of truncated CIPC scheme with $P_a^{\mathrm{max}}\rightarrow\infty$. However, we note the fact that the false alarm and miss detection rates at Willie for the conventional CIPC scheme cannot be directly obtained from that of the truncated CIPC scheme by setting $P_a^{\mathrm{max}} \rightarrow \infty$, due to the involved $\mathrm{Ei}$ functions. The complex $\mathrm{Ei}$ functions involved in the performance analysis on the truncated CIPC scheme leads to another fact that the impact of some system parameters (e.g., $P_b^{\max}$) cannot be clarified in the last section, even in the asymptotic scenario with $P_a^{\mathrm{max}}\rightarrow\infty$. These two facts motivates us to consider the conventional CIPC scheme and directly analyze its performance in this section, based on which we are able to analytically clarify the impact of some system parameters on the CIPC schemes and the comparison result between the truncated and conventional CIPC schemes.

\subsection{Detection Performance at Willie}

\setcounter{equation}{34}

In the conventional CIPC scheme, we assume that Alice will transmit a covert message with probability $1/2$, since Alice can always transmit covert information and the uninformative priors $\pi_0=\pi_1=1/2$ means a random guess of Alice's covert transmission at Willie. As such, the probability $\mathcal{P}(\Halt)$ and $\mathcal{P}(\Hnull)$ are, respectively, given by
\begin{align} \label{PH1}
\mathcal{P}(\Halt)&=\pi_1=\frac{1}{2},
\end{align}
and
\begin{align} \label{PH0}
\mathcal{P}(\Hnull)&=\pi_0=\frac{1}{2}.
\end{align}

Following the decision rule given in \eqref{decisions_FD}, we derive the false alarm and miss detection rates at Willie in the conventional CIPC scheme with the full-duplex Bob in the following theorem.
\begin{theorem}\label{theorem2}
The false alarm and miss detection rates at Willie are derived in \eqref{PFA_FD} and \eqref{PMD_FD}, respectively.
\begin{align}
\alpha&=\left\{
  \begin{array}{ll}
    1,  &\tau<\sigma_w^2, \\
    1-\frac{\tau-\sigma_w^2}{P_b^{\mathrm{max}}|h_{bw}|^2}, & \sigma_w^2\leq\tau\leq\nu,\\
    0,  &\tau>\nu,
  \end{array}
\right. \label{PFA_FD}
\end{align}
where we recall that $\nu$ is defined in \eqref{nu_definition}.
\end{theorem}

\setcounter{equation}{38}
\begin{IEEEproof}
Following \eqref{t_w_FD} and \eqref{decisions_FD}, the false alarm rate is given by
\begin{align}
\alpha&=\mathcal{P}\left[P_b|h_{bw}|^2+\sigma_w^2>\tau\right] \notag \\
&=\left\{
  \begin{array}{ll}
    1,  &\tau<\sigma_w^2, \\
    \int_{\frac{\tau-\sigma_w^2}{|h_{bw}|^2}}^{P_b^{\mathrm{max}}}f_{P_b}(x)\mathrm{d}x, &\sigma_w^2\leq\tau\leq\nu,\\
    0,  &\tau>\nu,
  \end{array}
\right.\notag \\
&=\left\{
  \begin{array}{ll}
    1,  &\tau<\sigma_w^2, \\
    \int_{\frac{\tau-\sigma_w^2}{|h_{bw}|^2}}^{P_b^{\mathrm{max}}}\frac{1}{P_b^{\mathrm{max}}}\mathrm{d}x, &\sigma_w^2\leq\tau\leq\nu,\\
    0,  &\tau>\nu.
  \end{array}
\right. \label{fa_proof2}
\end{align}
Then, solving the integral in \eqref{fa_proof2} leads to the desired result in \eqref{PFA_FD}.

As per \eqref{t_w_FD} and \eqref{decisions_FD}, the miss detection rate is given by
\begin{align}
\beta&=\mathcal{P}\left[P_a|h_{aw}|^2+P_b|h_{bw}|^2+\sigma_w^2<\tau\right] \notag \\
&=\left\{
\begin{array}{ll}
    0,  &\tau<\sigma_w^2, \\
    \mathcal{P}\left[P_b<\frac{\tau-\sigma_w^2-{Q|h_{aw}|^2}/{|h_{ab}|^2}}{|h_{bw}|^2}\right], &\sigma_w^2\leq\tau\leq\nu,\\
    \mathcal{P}\left[\frac{|h_{aw}|^2}{|h_{ab}|^2}<\frac{\tau-\sigma_w^2-P_b|h_{bw}|^2}{Q}\right],  &\tau>\nu,
\end{array}
\right. \label{md_proof3}
\end{align}
Following \eqref{md_proof3}, we achieve the desired result in \eqref{PMD_FD} after some algebra manipulations.
\end{IEEEproof}

We note that the false alarm and miss detection rates derived in Theorem~\ref{theorem2} are for an arbitrary detection threshold $\tau$. In practice, Willie will determine the optimal detection threshold that minimizes the total error probability, which is derived in the following theorem.

\begin{proposition}\label{proposition4}
For the decision rule given in \eqref{decisions_FD}, Willie's optimal threshold that minimizes the total error probability is derived as
\begin{align}
\tau^{\ast}=\nu,
\end{align}
and the corresponding minimum total error probability is given by
\begin{align} \label{xi ast}
\xi^{\ast}=1-\frac{Q\lambda_{aw}}{P_b^{\mathrm{max}} \lambda_{ab} |h_{bw}|^2}\ln\left(1+\frac{P_b^{\mathrm{max}} \lambda_{ab} |h_{bw}|^2}{Q\lambda_{aw}}\right).
\end{align}
where $\nu = P_b^{\mathrm{max}}|h_{bw}|^2+\sigma_w^2$ as defined in \eqref{nu_definition}.
\end{proposition}

\begin{IEEEproof}
As per \eqref{xi_definition}, \eqref{PFA_FD}, and \eqref{PMD_FD}, the total error probability at Willie is given by \eqref{xi_case}.

We first note that $\xi = 1$ is the worst case scenario for Willie and thus Willie does not set $\tau < \sigma_w^2$.
As per \eqref{xi_case}, for $\sigma_w^2 \leq \tau\leq\nu$ the total error probability $\xi$ is a monotonically decreasing function of $\tau$. Thus, Willie will set $\nu$ as the threshold to minimize $\xi$ in this case. For $\tau>\nu$, $\xi$ is an monotonically increasing function of $\tau$ based on \eqref{xi_case}, which indicates that Willie will try to set $\nu$ as the threshold again to minimize $\xi$ under this case. We also note that for the two cases, i.e., $\sigma_w^2 \leq \tau\leq\nu$ and $\tau>\nu$, setting $\tau = \nu$ can achieve the same total error probability. As such, we can conclude that the optimal detection threshold is $\tau$.
\end{IEEEproof}


\subsection{Optimization of Effective Covert Throughput}

In this subsection, we examine the maximum ECT achieved in the conventional CIPC scheme. The procedures for solving the optimization promblem are similar to that in Section~III.C.

\setcounter{equation}{43}

\begin{proposition}\label{proposition5}
The expected value of the minimum total error probability $\xi^{\ast}$ over all realizations of $h_{bw}$ is derived as
\begin{align} \label{xi ast_av}
\overline{\xi^{\ast}}(Q)
&\!=\!1\!-\!\frac{1}{4\varphi(Q)}\Bigg\{\!-\!\frac{8}{\varphi(Q)}{}_3F_3\left([1, 1, 1], [2, 2, 2], \frac{1}{\varphi(Q)}\right)\!+ \notag \\
&~~~4\varsigma\mathrm{Ei}\left(-\frac{1}{\varphi(Q)}\right)+4\ln(\varphi(Q))\Big(-\mathrm{Ei}\left(-\frac{1}{\varphi(Q)}\right)- \notag \\
&~~~\frac{\ln(\varphi(Q))}{2}+\varsigma\Big)+\pi^2-2\varsigma^2-2\Bigg\},
\end{align}
where $\varphi(Q)\triangleq P_b^{\mathrm{max}}\lambda_{ab}\lambda_{bw}/(Q\lambda_{aw})$, ${}_3F_3([\cdot, \cdot, \cdot],[\cdot, \cdot, \cdot],\cdot)$ is Gauss hypergeometric functions and $\varsigma$ is the Euler constant.
\end{proposition}

\begin{IEEEproof}
Following \eqref{xi ast}, the expected value of $\xi^{\ast}(|h_{bw}|^2,Q)$ with respect with $|h_{bw}|^2$ is given by
\begin{align}\label{expected_xi_proof}
\overline{\xi^{\ast}}(Q)&=\int_0^{\infty} \xi^{\ast}(|h_{bw}|^2,Q)\frac{e^{-\frac{|h_{bw}|^2}{\lambda_{bw}}}}{\lambda_{bw}} \mathrm{d} |h_{bw}|^2  \notag \\
&=1-\frac{Q\lambda_{aw}}{P_b^{\mathrm{max}} \lambda_{ab} \lambda_{bw}}\int_0^{\infty}\frac{e^{-\frac{|h_{bw}|^2}{\lambda_{bw}}}}{ |h_{bw}|^2}\times \notag \\
&~~~\ln\left(1+\frac{P_b^{\mathrm{max}} \lambda_{ab} |h_{bw}|^2}{Q\lambda_{aw}}\right) \mathrm{d} |h_{bw}|^2,
\end{align}
and solving the integral in \eqref{expected_xi_proof} with respect to $|h_{bw}|^2$ leads to the desired result in \eqref{xi ast_av}.
\end{IEEEproof}

The optimization problem at Alice of maximizing the ECT subject to a certain covert communication constraint is given by
\begin{align}\label{P1}
Q^{\ast} = &\argmax_{Q} R_c \\ \nonumber
&\text{s. t.} \quad  \min\{\pi_0, \pi_1\}\overline{\xi^{\ast}}(Q) \geq \min\{\pi_0, \pi_1\}- \epsilon.
\end{align}
where $\pi_0=\pi_1=1/2$ as defined in \eqref{PH0} and \eqref{PH1}, respectively. We note that the transmission outage probability of the conventional CIPC scheme is the same as that given in \eqref{outage_delta} of the truncated CIPC scheme and we have $\mathcal{P}_\mathbb{C}=1$ in the conventional CIPC scheme.

We next determine the solution to the optimization problem given in \eqref{P1} and derive the maximum ECT in the following theorem.

\begin{theorem}\label{theorem3}
For given $P_b^{\mathrm{max}}$ and $\epsilon$, the solution to \eqref{P1}, i.e., the optimal value of $Q$ that maximizes the ECT $R_c$ subject to $\overline{\xi^{\ast}}(Q) \geq 1-2\epsilon$, is derived as
\begin{align} \label{P_b_max_ast}
Q^{\ast}=Q_{\epsilon},
\end{align}
and the achieved maximum ECT is given by
\begin{align}\label{r_c}
R_{c}^{\ast}=R (1-e^{-\eta^{\ast}}-\eta^{\ast} \mathrm{Ei}(-\eta^{\ast})),
\end{align}
where $Q_{\epsilon}$ is the solution to $\overline{\xi^{\ast}}(Q)=1-2\epsilon$ of $Q$ and $\eta^{\ast}$ is given by
\begin{align}\label{eta_opt}
\eta^{\ast}=\frac{Q^{\ast}-(2^{R}-1)\sigma_b^2}{\left(2^{R}-1\right)\lambda_{bb}\phi P_b^{\mathrm{max}}}.
\end{align}
\end{theorem}

\begin{IEEEproof}
As per Corollary~\ref{corollary1}, the transmission outage probability $\delta$ is a monotonically decreasing function of $Q$, which leads to the fact that the ECT $R_c$ monotonically increases with $Q$. Following this fact, we next prove another fact that the expected minimum total error probability $\overline{\xi^{\ast}}(Q)$ is a monotonically increasing function of $Q$. These two facts indicate that the optimal value of $Q$ (i.e., the solution to \eqref{P1}) is the one that ensures $\overline{\xi^{\ast}}(Q) = 1 -2\epsilon$. To this end, we next prove that $\overline{\xi^{\ast}}(Q)$ is a monotonically decreasing function of $Q$.
Following \eqref{expected_xi_proof} and using the Leibniz integral rule, we derive the first derivative of $\overline{\xi^{\ast}}(Q)$ with respect to $t$ as
\begin{align}\label{proof1}
\frac{\partial \overline{\xi^{\ast}}(Q)}{\partial Q} &= \frac{\partial }{\partial Q}\left(\underbrace{\int_0^{\infty}\xi^{\ast}(|h_{bw}|^2,Q)\frac{e^{-\frac{|h_{bw}|^2}{\lambda_{bw}}}}{\lambda_{bw}}\mathrm{d}|h_{bw}|^2}_{\overline{\xi^{\ast}}(Q)}\right) \notag  \\
&=\int_0^{\infty}\frac{\partial \xi^{\ast}(|h_{bw}|^2,Q)}{\partial Q}\frac{e^{-\frac{|h_{bw}|^2}{\lambda_{bw}}}}{\lambda_{bw}}\mathrm{d}|h_{bw}|^2.
\end{align}
Considering ${e^{-\frac{x}{\lambda_{bw}}}}/{\lambda_{bw}} > 0$ in \eqref{proof1}, we could conclude that ${\partial \overline{\xi^{\ast}}(Q)}/{\partial Q} <0$ if we could prove $\xi^{\ast}(|h_{bw}|^2,Q)$ is a decreasing function of $Q$.
Following \eqref{xi ast}, we have
\begin{align}\label{proof2}
\frac{\partial \xi^{\ast}(\theta)}{\partial \theta}=\frac{u(\theta)}{\theta^2(1+\theta )},
\end{align}
where
\begin{align}
u(\theta) &\triangleq \left(1+\frac{\theta\lambda_{ab} |h_{bw}|^2}{\lambda_{aw}}\right)\ln\left(1+\frac{\theta\lambda_{ab} |h_{bw}|^2}{\lambda_{aw}}\right)- \notag  \\
&~~~~\frac{\theta\lambda_{ab} |h_{bw}|^2}{\lambda_{aw}}, \\
\theta&\triangleq \frac{P_b^{\mathrm{max}}}{Q}. \label{definition_theta}
\end{align}
We note that the sign of $\partial \xi^{\ast}(\theta)/\partial \theta$ depends on the value of $u(\theta)$. In order to determine the value range of $u(\theta)$, we first derive the first derivative of $u(\theta)$ with respect to $\theta$ as
\begin{align}\label{u_derivative}
\frac{\partial u(\theta)}{\partial \theta}=\frac{\lambda_{ab} |h_{bw}|^2}{\lambda_{aw}}\ln\left(1+\frac{\theta\lambda_{ab} |h_{bw}|^2}{\lambda_{aw}}\right) >0.
\end{align}
Noting $\lim_{\theta\rightarrow0}u(\theta) =0$ and following \eqref{u_derivative}, we can conclude that $u(\theta)\geq 0$ for $\theta>0$, which indicates that $\partial \xi^{\ast}(\theta)/\partial \theta \geq 0$. Noting that $\theta$ is a monotonically decreasing function of $Q$, we can conclude that $\xi^{\ast}(|h_{bw}|^2,Q)$ is a decreasing function of $Q$. This completes the proof of this theorem.
\end{IEEEproof}

\begin{corollary}\label{corollary2}
As Bob's maximum transmit power of the AN signal approaches infinity (i.e., $P_b^{\mathrm{max}} \rightarrow \infty$), the achievable maximum ECR approaches a fixed value, which is given by
\begin{align}\label{Rc_inf}
\lim_{P_b^{\mathrm{max}} \rightarrow \infty} R_c^{\ast}&=R\Bigg(1-\exp\left(-\frac{1}{(2^{R}-1)\lambda_{ab}\phi \theta_{\epsilon}}\right)+ \notag \\
&~~~\frac{\mathrm{Ei}\left(-\frac{1}{(2^{R}-1)\lambda_{ab}\phi \theta_{\epsilon}}\right)}{(2^{R}-1)\lambda_{ab}\phi \theta_{\epsilon}} \Bigg).
\end{align}
\end{corollary}
\begin{IEEEproof}
As per \eqref{xi ast} and \eqref{definition_theta}, we note that the solution of $\theta_{\epsilon}$ to $\overline{\xi^{\ast}}(\theta) = 1-2\epsilon$ does not affect by the value of $P_b^{\mathrm{max}}$. As such, we still have $t_{\epsilon}$ as the solution to $\overline{\xi^{\ast}}(\theta) = 1-2\epsilon$ as $P_b^{\mathrm{max}} \rightarrow \infty$. Then, substituting $Q^{\ast}=P_b^{\mathrm{max}}/\theta_{\epsilon}$ into \eqref{eta_opt} we have
\begin{align}\label{eta_opt_inf}
\lim_{P_b^{\mathrm{max}} \rightarrow \infty} \eta^{\ast} =\frac{1}{\left(2^{R}-1\right)\lambda_{bb}\phi \theta_{\epsilon}}.
\end{align}
Then, substituting \eqref{eta_opt_inf} into \eqref{r_c} we achieve the desired result in \eqref{Rc_inf} after some algebra manipulations.
\end{IEEEproof}

\section{Numerical Results}

In this section, we first present numerical results to verify our analysis on the covert communications in the considered two schemes (i.e., conventional CIPC and truncated CIPC). Then, we provide a thorough performance examination of the covert communications in each scheme. Based on our examination, we draw many useful insights and guidelines on how to design and implement covert communications in practical scenarios. Without other
statements, we set $\lambda_{ab}=\lambda_{aw}=\lambda_{bb}=\lambda_{bw}=1$.

\begin{figure} [t!]
\centering
\includegraphics[width=3.5in, height=2.9in]{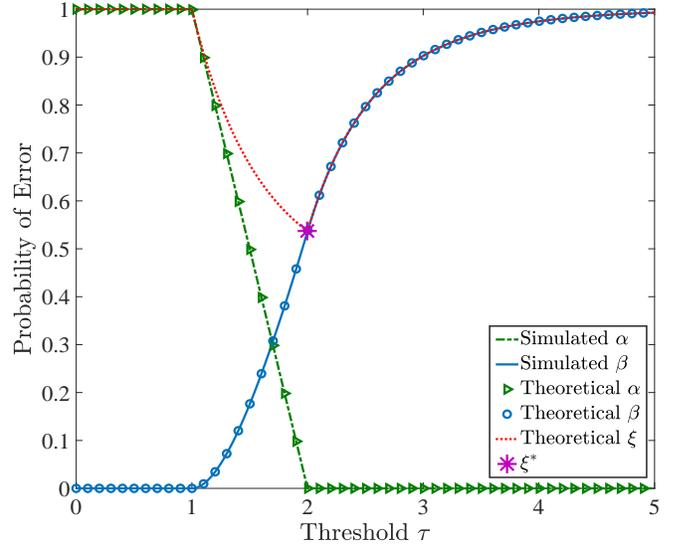}\\
\caption{Willie's false alarm rate $\alpha$, miss detection rate $\beta$, and total error probability $\xi$ versus threshold $\tau$ for covert communications in the truncated CIPC scheme, where $Q=1$, $P_b^{\mathrm{max}}=0$~dB, $P_a^{\mathrm{max}}=0$~dB, $|h_{bw}|^2=1$, and $\sigma^2_w=0$~dB.}\label{fig4}
\end{figure}

Fig.~\ref{fig4} illustrates the false alarm rate $\alpha$, miss detection rate $\beta$, and total error probability $\xi$ versus Willie's detection threshold $\tau$ in the truncated CIPC scheme. As expected, we first observe that the simulated curves precisely match the theoretical ones, which confirms the correctness of our Theorem~\ref{theorem2_trancated}.
We also observe that there is indeed an optimal value of $\tau$ that minimizes $\xi$ and this value satisfies $\tau^{\ast} =\nu$, which verifies the correctness of our Proposition~\ref{proposition4_trancated}.

\begin{figure} [t!]
\centering
\includegraphics[width=3.5in, height=2.9in]{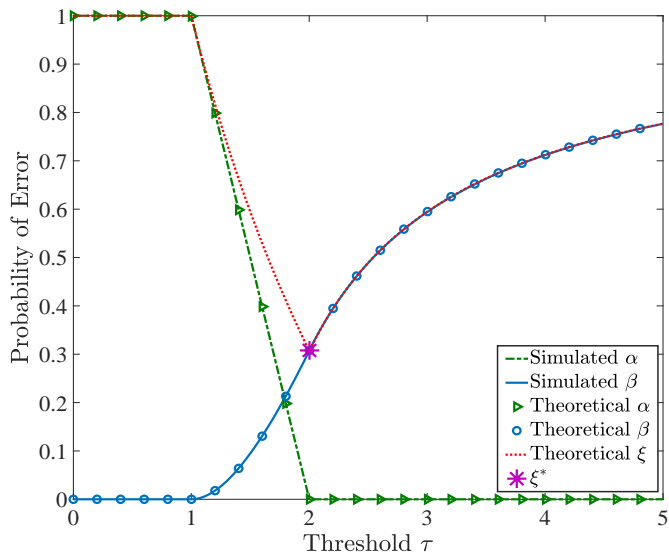}\\
\caption{Willie's false alarm rate $\alpha$, miss detection rate $\beta$, and total error probability $\xi$ versus threshold $\tau$ for covert communications in the conventional CIPC scheme, where $Q=1$, $P_b^{\mathrm{max}}=0$~dB, $|h_{bw}|^2=1$, and $\sigma^2_w=0$~dB.}\label{fig3}
\end{figure}

In Fig.~\ref{fig3}, we plot the false alarm rate $\alpha$, miss detection rate $\beta$, and total error probability $\xi$ versus Willie's detection threshold $\tau$ in the conventional CIPC scheme.
As expected, in this figure we first observe that the simulated curves precisely match the theoretical ones, which confirms the correctness of our Theorem~\ref{theorem2}. We also observe that the total error probability $\xi$ dramatically varies with the detection threshold $\tau$, which demonstrates the necessity of optimizing $\tau$ by Willie and the importance of our Proposition~\ref{proposition4}, which derives the optimal detection threshold minimizes $\xi$ in a closed-form expression. Finally, we observe that the optimal value of $\tau$ is equal to $\nu$, which simultaneously forces the false alarm rate $\alpha$ at Willie being zero and minimizes the miss detection rate $\beta$.

\begin{figure} [t!]
    \begin{center}
    \includegraphics[width=3.5in, height=2.9in]{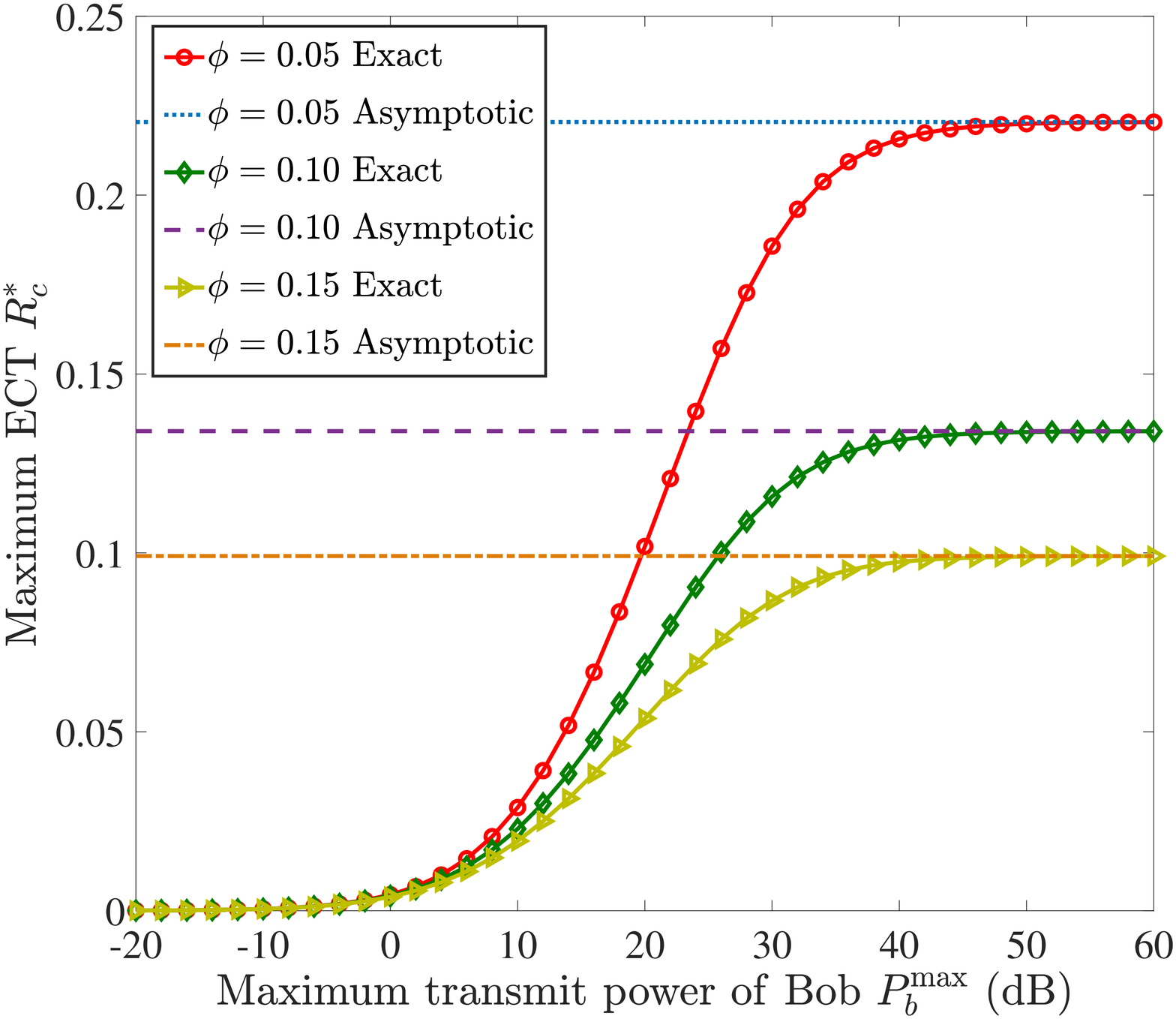}
    \caption{Maximum ECT $R_c^{\ast}$ versus Bob's maximum transmit power $P_b^{\mathrm{max}}$ under different value of $\phi$ in the conventional CIPC scheme, where $\epsilon=0.10$ and $\sigma^2_b=0$~dB.}\label{fig5}
    \end{center}
\end{figure}

In Fig.~\ref{fig5}, we plot the maximum ECT $R_c^{\ast}$ versus Bob's maximum transmit power of the AN signal $P_b^{\mathrm{max}}$ with different values of the self-interference cancellation parameter $\phi$. In this figure, we first observe that $R_c^{\ast}$ monotonically increases as $P_b^{\mathrm{max}}$ increases, which demonstrates that the covert communications from Alice becomes easier when Bob has more power to transmit AN to aid. However, we also note that as $P_b^{\mathrm{max}} \rightarrow \infty$ the maximum ECT $R_c^{\ast}$ approaches the upper bound given in our Corollary~\ref{corollary2}. Intuitively, this can be explained by the fact that the transmitted AN not only creates interference at Willie but also leads to self-interference at Bob. In this figure, we also observe that the upper bound on the achieved $R_c^{\ast}$ decreases significantly as $\phi$ increases, since a larger $\phi$ indicates a larger self-interference at Bob.

\begin{figure} [t!]
    \begin{center}
    \includegraphics[width=3.5in, height=2.9in]{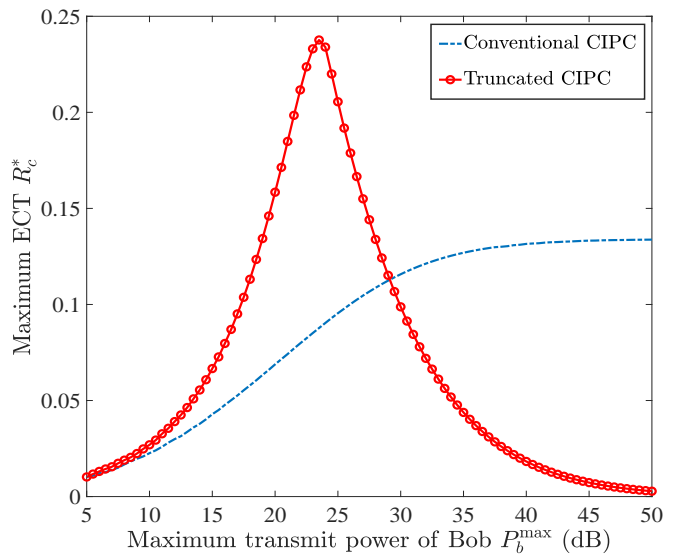}
    \caption{Maximum ECT $R_c^{\ast}$ versus Bob's maximum transmit power $P_b^{\mathrm{max}}$ optimized over $Q$ and $R$, where $\phi=0.1$, $\epsilon=0.01$, $P_a^{\mathrm{max}}=10$~dB, and $\sigma^2_b=0$~dB.}\label{fig6}
    \end{center}
\end{figure}

In Fig.~\ref{fig6}, we plot the maximum ECT $R_c^{\ast}$, which is achieved by the optimal $Q$ and $R$, versus Bob's maximum transmit power of the AN signal $P_b^{\mathrm{max}}$. Since $R$ has no effect on the detection performance of Willie, we can obtain the optimal $R$ by a numerical search, after obtaining the optimal $Q$. In this figure, we first observe that $R_c^{\ast}$ achieved by the truncated CIPC scheme first increases and then decreases as $P_b^{\mathrm{max}}$ increases, which indicates that there is an optimal value of $P_b^{\mathrm{max}}$ that maximizes $R_c^{\ast}$. This due to the fact a small $P_b^{\mathrm{max}}$ cannot cause enough uncertainty at Willie to hide Alice's covert transmission, while a large $P_b^{\mathrm{max}}$ may have greater impact on increasing the interference at Bob than increasing uncertainty at Willie (since $P_a^{\mathrm{max}}$ is predefined in the truncated CIPC scheme).
Meanwhile, we also observe that the $R_c^{\ast}$ achieved by the conventional CIPC scheme continuously increases as $P_b^{\mathrm{max}}$ increases, which is due to the fact that in the conventional CIPC scheme we can vary $Q$ to counteract the impact of $P_b^{\mathrm{max}}$ on the detection performance at Willie (can be seen from Corollary~\ref{corollary2}) and the outage probability converges to a specific constant value as $P_b^{\mathrm{max}}$ increases (can be seen from \eqref{eta_opt_inf}). Furthermore, in this figure we observe that the maximum value of $R_c^{\ast}$ for the truncated CIPC scheme is higher than that for the conventional CIPC scheme, which demonstrates that the truncated CIPC scheme can outperform the conventional CIPC scheme. Intuitively, this is due to the fact that increasing $P_a^{\mathrm{max}}$ can decrease the detection error probability at Willie (i.e., it becomes easier for Willie to detection Alice's covert transmission), although increasing $P_a^{\mathrm{max}}$ always increases the effective rate from Alice to Bob.

\begin{figure} [t!]
    \begin{center}
    \includegraphics[width=3.5in, height=2.9in]{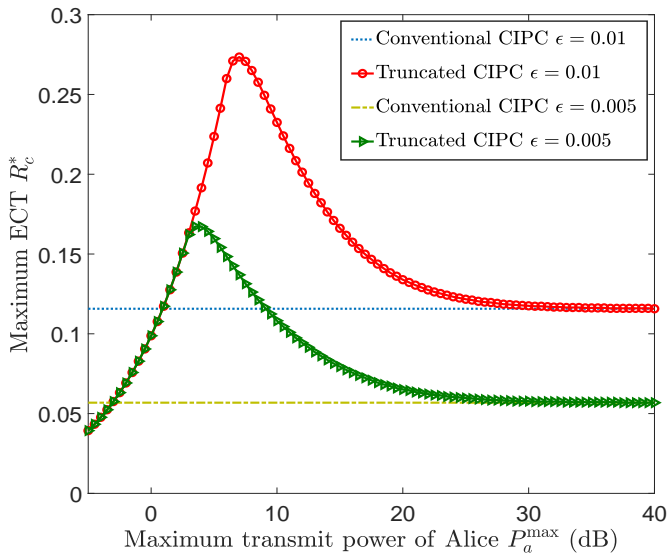}
    \caption{Maximum ECT $R_c^{\ast}$ versus Alice's maximum transmit power $P_a^{\mathrm{max}}$ optimized over $Q$ and $R$ under different value of $\epsilon$, where $P_b^{\mathrm{max}}=20$~dB, $\sigma^2_b=-10$~dB, and $\phi=0.1$.}\label{fig7}
    \end{center}
\end{figure}

In Fig.~\ref{fig7}, we plot the maximum ECT $R_c^{\ast}$, achieved by the conventional and truncated CIPC schemes with the optimal $Q$ and $R$, versus the maximum transmit power at Alice $P_a^{\mathrm{max}}$ with different levels of covert communication constraints (i.e., different values of $\epsilon$). In this figure, we first observe that $R_c^{\ast}$ achieved by the truncated CIPC scheme first increases and then decreases as $P_a^{\mathrm{max}}$ increases, which indicates that there is an optimal value of $P_a^{\mathrm{max}}$ that maximizes $R_c^{\ast}$. This can be explained by the fact that increasing $P_a^{\mathrm{max}}$ simultaneously decreases the detection error probability at Willie and the probability that Alice can transmit covert information to Bob (i.e., $\mathcal{P}_\mathbb{C}$), which means that $P_a^{\mathrm{max}}$ has a two-side impact on the considered covert communications. This observation indicates, on top of optimizing the fixed value $Q$, we also have to optimally design the parameter $P_a^{\mathrm{max}}$ in order to achieve the optimal performance of CIPC in the context of covert communications. This is different from optimizing the performance of CIPC in point-to-point communications, where the performance of CIPC monotonically increases with $P_a^{\mathrm{max}}$ and thus solely optimizing $Q$ is sufficient. We also observe that, as $P_a^{\mathrm{max}}$ increases, the $R_c^{\ast}$ of the truncated CIPC scheme approaches that of the conventional CIPC scheme, which confirms the correctness of our examinations. As expected, we also observe that $R_c^{\ast}$ increases as $\epsilon$ increases, which demonstrates that it is the covert communication constraint that mainly limits $R_c^{\ast}$.


\section{Conclusion}

This work examined covert communications with a full-duplex receiver over Rayleigh fading channels, in which the transmitter Alice adopts the CIPC to transmit information to the receiver Bob covertly while trying to hide herself from the warden Willie. We analyzed Willie's detection performance limits for the considered truncated and conventional CIPC schemes, based on which the achievable ECTs of these two schemes were determined. Our examination shows that there exists an optimal value of the maximum transmit power at Bob that maximizes the ECT in the truncated CIPC scheme, while the ECT monotonically increases and approaches an upper bound as this maximum transmit power at Bob increases in the conventional CIPC scheme. Our analysis and examinations provided practical guidelines on conducting covert communications and potentially hiding a transmitter in Rayleigh fading scenarios.


\bibliographystyle{IEEEtran}
\bibliography{IEEEfull,CC}

\end{document}